# A Disciplined Approach to Adopting Agile Practices: The Agile Adoption Framework

Ahmed Sidky, James Arthur
(asidky@vt.edu, arthur@vt.edu)
*Virginia Tech*

## Abstract

*Many organizations aspire to adopt agile processes to take advantage of the numerous benefits that it offers to an organization. Those benefits include, but are not limited to, quicker return on investment, better software quality, and higher customer satisfaction. To date however, there is no structured process (at least in the public domain) that guides organizations in adopting agile practices. To address this problem we present the Agile Adoption Framework. The framework consists of two components: an agile measurement index, and a 4-Stage process, that together guide and assist the agile adoption efforts of organizations. More specifically, the agile measurement index is used to identify the agile potential of projects and organizations. The 4-Stage process, on the other hand, helps determine (a) whether or not organizations are ready for agile adoption, and (b) guided by their potential, what set of agile practices can and should be introduced.*

## 1. Introduction and Motivation

Over the past few years organizations have asked the agile community *"Why should we adopt agile practices?"* [27]. The numerous success stories highlighting the benefits reaped by organizations that have successfully adopted agile practices provide an answer to this question [49] [41] [9] [8] [34] [32]. As a result, many organizations are now aspiring to adopt agile practices. Once again, however, they are turning to the agile community, but with a different question: *"How do we proceed with adopting agile practices?"* [27]. Unfortunately, there exists no structured approach (at least in the public domain) for agile adoption. The absence of guidance and assistance to organizations pursuing agility is the main problem addressed by this paper.

A major factor contributing to this absence is the number of issues a structured approach must address when providing organizations with guidance for the successful adoption of agile practices. These include, among other issues, determining: (1) the organization's readiness for agility; (2) the practices it should adopt; (3) the potential difficulties in adopting them; (4) and finally, the necessary organizational preparations for the adoption of agile practices.

The Agile Adoption Framework introduced in this paper, is an attempt to addresses the issues mentioned above by providing a structured and repeatable approach designed to guide and assist agile adoption efforts. It assists the agile community in supporting the growing demand from organizations that want to adopt agile practices. The Agile Adoption Framework, however, is only one essential ingredient, the other is an agile coach who knows how to apply that framework. Such a person can be an agile consultant hired to facilitate the process, or an in-house employee with sufficient training on agile methods and the use of the framework.

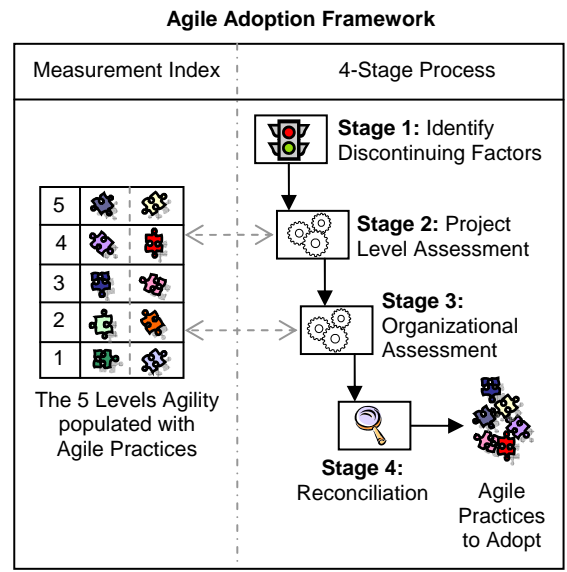

**Figure 1. Overview of the Agile Adoption Framework**

The Agile Adoption Framework has two main components: (1) a measurement index for estimating agile potential, and (2) a 4-Stage process that employs the measurement index in determining which, and to what extent, agile practices can be introduced into an organization. Figure 1 illustrates the various components of the framework and the relationships among them.

The first component, the agile measurement index, is a scale the coach uses to identify the agile potential of a project or organization. The agile measurement index is used in the process component of the framework, which consists of four stages working together to guide organizations in identifying agile practices that best fit into their environment. The four stages are:

- *Stage 1: Identification of Discontinuing Factors.* Discovers the presence of any showstoppers that can prevent the adoption process from succeeding.

- *Stage 2: Project Level Assessment.* Utilizes the agile measurement index to determine the target level of agility for a particular project.

- *Stage 3: Organizational Readiness Assessment.* Uses the agile measurement index to assess the extent to which the organization can achieve the target agility level identified for a project.

- *Stage 4: Reconciliation.* Determines the final set of agile practices to be adopted by reconciling the target agile level for a project (from Stage 2) with the readiness of the embodying organization (from Stage 3).

Section 2 presents the structure and details of the agile measurement index. Each of the four stages in the process is then presented in detail in Section 3. Section 4 presents industry feedback regarding the framework. Section 5 provides concluding remarks about the Agile Adoption Framework along with comments from the agile community.

## 2. Agile Measurement Index

One of the concerns organizations have when seeking to adopt agile practices is determining how agile they can become [23]. The agile potential (i.e. the degree to which that entity can adopt agile practices) of projects and organizations is influenced by the circumstances surrounding them. To determine the agile potential the coach (or the one conducting the assessment) needs use a measurement index or scale that can assess the agility of an entity. The agile adoption framework refers to this scale as an *agile measurement index*.

The Agile Adoption Framework uses the agile measurement index to determine the agile potential of projects and organizations. The measurement index is composed of four components:

1. Agile Levels
2. Agile Principles
3. Agile Practices and Concepts
4. Indicators.

Sections 2.1 through 2.4 introduce each component of the agile measurement index. Section 2.5 focuses on issues related to the tailorability of the index.

### 2.1. Agile Levels

*Agile levels*, as depicted in Figure 2a, are considered the units of the measurement scale as they enumerate the different possible degrees of agility for a project or organization. The agile potential of a project or organization is expressed in terms the highest agile level it can achieve. The attainment of a particular level symbolizes that the project or organization has realized and embraced the essential elements needed to establish a particular degree of agile effectiveness. For example, when the elements inherent to *enhancing communication and collaboration* are embodied within the development process, then the Agile Level 1 (*Collaborative*) is attainted. However, before one can expect to move to Level 2 status, all practices associated with Agile Level 1 must be achieved (or achievable).

The 5 Levels of Agility are designed to represent the core qualities of the Agile Manifesto [2], rather than the qualities related to any particular agile method. After careful analysis of the manifesto, five essential agile qualities have been identified. Those qualities comprise the 5 Levels of Agility that are used the agile measurement index:

- *Level 1: Collaborative.* This level denotes the fostering of *communication and collaboration between all stakeholders*. The dimension of collaboration is the foundation of agile software development [45] [17] [18].

- *Level 2: Evolutionary.* Evolutionary development is the *early and continuous delivery of software*. It too is fundamental because every agile method assumes its presence [33].

- *Level 3: Effective.* The next quality an agile development process must embrace is that of *developing high quality, working software in an efficient an effective manner*. This quality is needed to prepare the development process so that it can respond to constant change without jeopardizing the software system being developed [29] [18].

- *Level 4: Adaptive.* This level constitutes establishing the agile quality of *responding to change in the process*. Defining and responding to multiple levels of feedback is essential in this level [26].

- *Level 5: Ambient.* The last level concentrates on establishing a *vibrant environment* needed to sustain and foster agility throughout an organization.

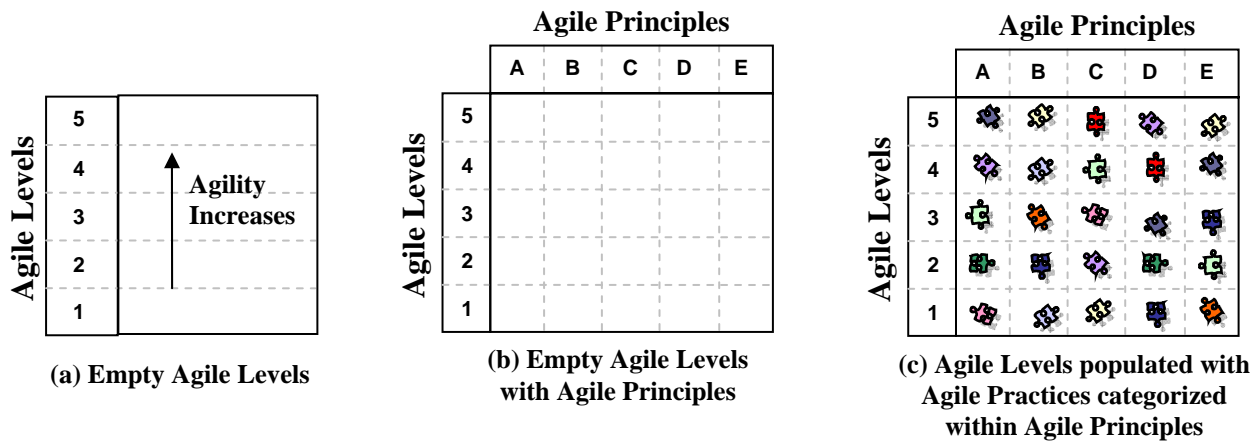

**Figure 2. Components of the 5 Levels of Agility (Indicators are not shown)**

Each of the agile levels is composed of a set of agile practices that introduce and sustain the agile quality pertinent to that level. The selection of agile practices and concepts assigned to each agile level is guided by the second component of the measurement index, *agile principles*.

## 2.2. Agile Principles

Agile principles are the essential characteristics that must be reflected in a process before it is considered *Agile*. For example, two key agile principles are *human centric*, which refers to the reliance on people and the interaction between them, and *technical excellence*, which implies the use of procedures that produce and maintain the highest quality of code possible. The Agile Manifesto outlines 12 principles that characterize agile development processes [13]. After careful grouping and summarization, five agile principles emerge that capture the essence of the 12. These five principles guide the refinement or tailoring of the 5 Levels of Agility:

- *Embrace change to deliver customer value* [12]. The success of a software development effort is based on the extent to which it helps deliver customer value. In many cases the development team, as well as the customer, are in a continuous learning process as to the requirements necessary to realize additional customer value. Hence, an attitude of welcoming and embracing change should be maintained throughout the software development effort.

- *Plan and deliver software frequently* [13] [20] [38]. Early and frequent delivery of working software is crucial, because it provides the customer with a functional piece of the product to review and provide feedback on. This feedback is essential for the process of planning for upcoming iterations as it shapes the scope and direction of the software development effort.

- *Human centric* [17]. The reliance on people and the interactions among them is a cornerstone in the definition of agile software processes.

- *Technical excellence* [26] [31]. Agile developers are committed to producing only the highest quality code possible, because high quality code is essential in fast-paced development environments, such as the ones characterized as agile.

- *Customer collaboration* [13]. Inspired from the original statement of the agile manifesto, there must be significant and frequent interaction between the customers, developers, and all the stakeholders of the project to ensure that the product being developed satisfies the business needs of the customer.

In effect, agile principles are used to ensure that the agile levels embody the essential characteristics of agility. Figure 2b illustrates the relationship between agile levels and agile principles. Each agile level should contain agile practices associated with most, if not all, of the agile principles. The principle reflects the approach that the agile practice uses to promote the agile quality pertinent to a level. For example, all the practices in Level 3 (Effective) are promoting the agile objective of *developing high quality, working software in an efficient an effective manner. How* that objective is achieved though, is determined by the practices associated with agile principles spanning each level. Along the same lines, practices associated with the *technical excellence* principle will promote its agile objective by focusing on enhancing the technical aspect of the process, while practices associated with the *human centric* principle promote enhancing the human aspect of the process.

|  | **Agile Principles** | | | | |
| --- | --- | --- | --- | --- | --- |
|  | *Embrace Change to Deliver Customer Value* | *Plan and Deliver Software Frequently* | *Human Centric* | *Technical Excellence* | *Customer Collaboration* |
| Level 5 **Ambient** *Establishing a vibrant environment to sustain agility* | Low process ceremony [33, 38] | Agile project estimation [20] | Ideal agile physical setup [33] | Test driven development [11]<br><br>Paired programming [48]<br><br>No/minimal number of level -1 or 1b people on team [17, 15] | Frequent face-to-face interaction between developers & users (collocated) [12] |
| Level 4 **Adaptive** *Responding to change through multiple levels of feedback* | Client driven iterations [33]<br><br>Continuous customer satisfaction feedback [35, 42] | Smaller and more frequent releases (*4-8 weeks*) [35]<br><br>Adaptive planning [33] [20] |  | Daily progress tracking meetings [6]<br><br>Agile documentation [39, 31]<br><br>User stories [21] | Customer immediately accessible [15]<br><br>Customer contract revolves around commitment of collaboration [26, 35] |
| Level 3: **Effective** *Developing high quality, working software in an efficient an effective manner* |  | Risk driven iterations [33]<br><br>Plan features not tasks. [20]<br><br>Maintain a list of all features and their status (backlog) [31] | Self organizing teams [33, 38, 31, 18]<br><br>Frequent face-to-face communication [38, 18, 13] | Continuous integration [33]<br><br>Continuous improvement (refactoring) [31, 12, 24, 5].<br><br>Unit tests [28]<br><br>30% of level 2 and level 3 people [17, 15] |  |
| Level 2: **Evolutionary** *Delivering software early and continuously* | Evolutionary requirements [33] | Continuous delivery [33, 31, 26, 12]<br><br>Planning at different levels [20] |  | Software configuration management [31]<br><br>Tracking iteration progress [33]<br><br>No big design up front (BDUF) [4, 12] | Customer contract reflective of evolutionary development [26, 35] |
| Level 1: **Collaborative** *Enhancing communication and collaboration* | Reflect and tune process [35, 42] | Collaborative planning [38, 18, 33] | Collaborative teams [45]<br><br>Empowered and motivated teams [13] | Coding standards [29, 47, 36]<br><br>Knowledge sharing tools [33]<br><br>Task volunteering [33] | Customer commitment to work with developing team [13] |

**Table 1. The 5 Levels of Agility populated with Agile Practices and Concepts**

The real essence of the 5 Levels of Agility, however, is in the agile practices it enunciates. The next section presents the third component of the agile measurement index – the *agile practices.*

### 2.3. Agile Practices

Agile practices are concrete activities and practical techniques that are used to develop and manage software projects in a manner consistent with the agile principles. For example, *paired programming*, *user stories,* and *collaborative planning* are all agile practices. Since the agile levels are composed of agile practices (organized along the line of agile principles – see Figure 2c), they are considered the basic building block of the agile measurement index. The attainment of an agile level is achieved only when the agile practices associated with it are adopted.

After surveying the agile methods currently used in industry [29] [31] [3], 40 distinct agile practices were chosen to populate the 5 Levels of Agility. These practices, arranged along the lines of the agile levels and principles, are illustrated in Table 1. (The underlining of the practices should be ignored at this point, but is discussed later in the paper.) Although a detailed discussion about each of the agile practices and concepts is outside the scope of this paper, the references associated with each are good starting points to learn more about them.

## 2.4. Indicators

A set of indicators, or questions, must accompany each agile practice or concept in the measurement index. The agile coach uses these indicators (or questions) to measure the extent to which the organization is ready to adopt an agile practice or concept. The Goal Question Metric approach (GQM) [10] and the Objectives Principles Attributes (OPA) Framework [7] influence the approach used to devise the indicators for each practice. Each indicator is designed to measure a particular organizational characteristic necessary for the successful adoption of the agile practice to which the indicator is related. (This is the goal.) Depending on the question, a manager, developer, or the agile coach is designated to answer it, either subjectively or objectively.

For example, assume the coach wants to determine the extent to which an organization is ready to adopt *coding standards* (Level 1, Technical Excellence). In this respect, two organizational characteristics that need to be assessed are: (1) to what extent do the developers understand the benefits behind coding standards, and (2) how willing are they to conform to coding standards. Several indicators (or questions) are used to assess each of these characteristics. For example, to assess the second (willingness), the assessor might ask the developers *to what extent would they abide by coding standards even when under a time constraint.*

The 5 Levels of Agility contain approximately 300 different indicators for the 40 agile practices. A detailed listing of all the indicators associated with each agile level is found in the framework's technical documentation [43].

The 5 Levels of Agility shown in Table 1 is one instance of the agile measurement index. Can there be, however, alternative instances? We address that issue in the next section.

## 2.5. Tailorability of the 5 Levels of Agility

The 5 Levels of Agility, along with all their practices and indicators, were presented to members of the agile community. Several of its leaders encouraged us to consider factors that might lead to other instances of the 5 Levels of Agility. These factors are *incorporating business values* and *reorganizing the practices based on experiential success*. The two following subsections elaborate on these factors.

**2.5.1 Incorporating Business Values.** Business values refer to the added benefit realized by an organization after adopting agile practices. For most organizations, the achievement of these business values is the real incentive behind adopting agility. For example, *decreasing time to market* or *increasing product quality* are common business values that organizations hope to realize from adopting agile practices. Augustine [40] and Elssamadisy [22] have suggested that the levels of agility might be prioritized according to the business values an organization hopes to realize. This suggestion is both valuable and beneficial to the growth of the framework, because currently, the 5 Levels of Agility are not associated with any business values; instead they are based on the qualities and values of agility. The relationship between agile and business values is parallel to that between the Agile Manifesto (focusing on agile values) and the Declaration of Interdependence (capturing the business values) [2] [1].

**2.5.2 Reorganizing the Practices based on Experiential Success.** The agile coaches and consultants Cockburn [16], Cohn [19], and Wake [46], in addition to others, suggest a reorganization of the agile practices based on experiential successes. That is, they advocate that the type of project and the experiences gained from previous adoption efforts can, and should, serve as a basis for formulating a better arrangement of the practices within the agile levels. For example, Cohn suggests that *user stories* be introduced in the first level of agility, because, from his experience, they enhance collaboration and communication between the stakeholders with regards to requirements. Others suggest that *pair programming* be in the first level because it helps establish collaboration within teams.

This inability to reach a consensus on the position of agile practice emphasizes an important factor in providing guidance in an agile adoption effort: the *adherence* to agile principles when establishing the levels is paramount, not the *positions* of the actual practices. The intention behind the levels of agility is to provide a framework to guide the adoption process, not to dictate it

Based on the above rationalizations we must conclude that a tailorable measurement index is both desirable and beneficial. However, when tailoring or creating another instance of an agile measurement index, it is important to observe the following guidelines to ensure that the new measurement index has all the necessary components and a valid structure:

- *Ensure that multiple levels exist*. Levels are needed to enumerate the degrees of agility. Without levels, the power of the measurement index, when used in conducting comparative measurements of agility, is diminished.

- *The measurement index is based on practices and concepts*. Foundational to the agile measurement index are agile practices and concepts. The extent to which agile practices and concepts can be adopted determines the agility of a process.

- *Each practice or concept has indicators*. When introducing a new agile practice (other than the 40 identified) to the measurement index, it is important that the practice has an associated set of valid and sufficient indicators. Without indicators, there is no means by which an assessment can be conducted.

The next section presents the second component of the Agile Adoption Framework – the 4-Stage Process. This component utilizes the 5 Levels of Agility (i.e., the agile measurement index) to provide structured guidance and assistance to organizations seeking to adopt agile practices.

## 3. The 4-Stage Process for Agile Adoption

The 4-Stage assessment process is the "backbone" of the Agile Adoption Framework. As depicted in Figure 3, it first provides an assessment component that helps determine if (or when) an organization is ready to move toward agility, i.e., make the go/no-go decision. Secondly, the process guides and assists the agile coach in the process of identifying which agile practices the organization should adopt. The four stages are grouped according to the objective they help to achieve:

- Objective 1: Make Go/No-Go Decision
  o *Stage 1: Discontinuing Factors*
- Objective 2: Identify Agile Practices to Adopt
  o *Stage 2: Project Level Assessment*
  o *Stage 3: Organizational Readiness Assessment*
  o *Stage 4: Reconciliation*

The next sections explain in detail how each stage of the 4-Stage process contributes to achieving its enunciated objectives.

### 3.1. Making the Go/No-Go Decision

The first objective of the process is to provide organizations with a method for deciding whether or not to proceed with agile adoption initiatives. Since adopting agile practices is essentially a type of Software Process Improvement (SPI), a pre-assessment phase is needed before the decision to start the initiative is made. Traditionally, pre-assessments determine the ability of the organization to undertake an SPI initiative [25]. Organizations lacking the factors necessary for a successful SPI effort are considered "not ready." In that situation the SPI effort is suspended until the missing factors can be mitigated.

Similarly, with respect to agile adoption, pre-assessment helps identify factors in an organization that can prevent the successful adoption of agile practices. If such factors exist, the organization must eliminate them before continuing with the adoption effort. Pre-assessment processes like these are important because they save the organization time, money and effort by identifying missing or existing factors that can cause an SPI initiative to fail [30].

The next section describes how Stage 1 of the process guides and assists organizations in making Go/No-go decisions concerning the adoption of agile practices. This decision is determined by a pre-assessment activity that identifies any discontinuing factors.

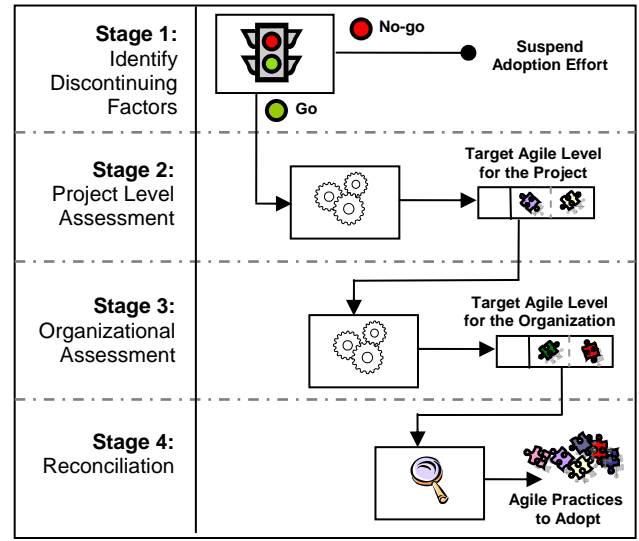

**Figure 3. The 4-Stage Process for Agile Adoption**

**3.1.1 Stage 1: Identifying Discontinuing Factors.** The intent of Stage 1 is to provide an assessment process that identifies factors which could prevent the successful adoption of agile practices. These are called *discontinuing factors,* and can vary from one organization to another. Typically, they pertain to an organization's resources including money, time and effort, as well as the support of its executive leadership. The three discontinuing factors identified by the Agile Adoption Framework are:

- *Inappropriate Need for Agility:* This refers to situations where, from a business or software development perspective, adopting agility does not add any value [44].

- *Lack of Sufficient Funds:* When funds are unavailable or insufficient to support the agile adoption effort, then an adoption process is not feasible.

- *Absence of Executive Support:* If committed support from executive sponsors is absent, then effective and substantial change in the organization is unlikely to occur [44] [37].

When an organization demonstrates *any* of these discontinuing factors, it is unprepared to move towards agility and should suspend the adoption process until the environment is more supportive.

Indicators focusing on organizational characteristics are used to assess the degree to which a discontinuing factor is present in the organization. The assessor uses one or more indicators to evaluate each organizational characteristic. For example, two organizational characteristics that can be measured to determine whether there is a *Lack of Sufficient Funds* are (1) the dollar amount allocated to the process improvement effort and (2) the ability to actually spend the funds for agile adoption. An example of a question (indicator) used to assess the ability to spend funds on agile adoption is *Can the funds be spent towards any process improvement activity?* Another assessment question is *Are there any restrictions on the type of activities for which these funds can be used?* Over 20 indicators are included in the Agile Adoption Framework to assess the presence of discontinuing factors in organizations [43].

## 3.2. Identify Agile Practices to Adopt

If Stage 1 indicates that the organization is ready to move towards agility, the journey of introducing agile practices into the development process begins. This involves determining which agile practices and concepts are most suitable for the organization to adopt. Actually, to be more precise, the Agile Adoption Framework first determines the agile practices that a *particular project* can adopt, not the whole organization. The framework is based on the fundamental belief that each project in an organization can adopt a different degree of agility based on its context. Therefore, the last three stages provide guidelines for identifying the agile practices suitable for a single project:

- *Stage 2: Project level Assessment:* identifies the maximum level of agility the project can reach. This is also known as the target agile level.
- *Stage 3: Organizational Readiness Assessment:* determines the extent to which the organization is ready to accommodate the project's target agile level.
- *Stage 4: Reconciliation:* settles the differences, if any, between the highest level of agility the project can adopt and the level of agility the organization is ready to embrace, and determines the agile practices that are to be adopted.

Sections 3.2.1 through 3.2.3 describe each of these stages, respectively.

**3.2.1 Stage 2: Project level Assessment.** Stage 2 is the first stage of the adoption process that utilizes the 5 Levels of Agility presented earlier. The objective of this stage is to identify the highest level of agility a project can achieve. This is called the target level, and is one of the 5 agile levels.

In theory, all projects should aspire to reach the highest level of agility possible. However, the reality is that circumstances, often outside of the organization's control, surround each project. These circumstances become constraining factors if they adversely affect the organizations' ability to adopt an agile practice. Thus, constraining factors limit the level of agility to which a project aspires.

For example, *frequent face-to-face communication* is a desired agile practice at level 3. A factor that is needed to successfully adopt this practice is *near team proximity*. Assume that the project and organization have no say in changing this project characteristic (i.e. factor), because it is outside of their control. If the project level assessment determines that the factor (near team proximity) is missing for this project, then the highest level of agility for this project will be the same level of agility in which this agile practice is found (which is Level 3 in this case).

Because achieving the highest level of agility depends on project circumstances outside of an organization's control, the first step in Project Level Assessment is to identify those agile practices and concepts that rely on such circumstances. These agile practices are known as *limiting agile practices*, because if the project characteristics needed to support these practices are not present, the inability to adopt the practice constrains or limits the level of agility attainable by the project. In Table 1, which illustrates the 5 agile levels, the limiting agile practices are underlined.

The assessment process defined by Stage 2 focuses on determining the target level of agility for a project. More specifically, it examines only those factors associated with the limiting agile practice, and measures the extent to which they are present. The assessment is conducted using the indicators associated with each limiting agile practice. The process starts by examining the limiting practices at Agile Level 1, and then moves upward on the scale. Once factors needed for the adoption of a limiting practice are found to be missing, the assessment process stops, and the highest level of agility attainable for the project is set to be the level at which that limiting practice is found.

In summary, the target level of agility is determined to be the point where the assessment process discovers that one of the project characteristics needed to adopt a limiting agile practice or concept is missing, and neither the project nor organization can do anything to influence or change this circumstance. After the target agile level for the project is identified, the next step in the journey is to conduct an organizational readiness assessment to determine the set of agile practices (for the project) that *can* be adopted.

**3.2.2 Stage 3: Organizational Readiness Assessment.** Identifying the target level for a project does not necessarily mean that that level is *achievable*. To determine the extent to which that target level can be achieved, the organization must be assessed to determine whether it is ready to adopt each of the agile practices and concepts associated up to, and including, the target level. Investing time and effort in this type of pre-adoption assessment of each agile practice increases the probability of success for the overall transition to agility [14], because it significantly reduces the risks associated with the agile adoption process.

Similar to Stage 2, Stage 3 of the process also relies on the 5 Levels of Agility. Again, the indicators play a critical role in determining the extent to which the target level can be achieved. To save time and money during this assessment stage, instead of assessing how ready the organization is relative to adopting the practices in all 5 agile levels, only those within the target agile level and below are used. The assessor uses the set of indicators (questions) associated with the agile practices to measure the extent to which the requisite organizational characteristics are present.

For example, *Collaborative Planning* is an agile concept in Level 1. To assess the readiness of the organization to adopt this concept, the following are some of the organizational characteristics that need to be present: (a) collaborative management style, (b) management buy-in to adopt the agile practice, (c) transparency of management, (d) small power-distance in the organization, and (e) developers buy-in to adopt the agile practice

| Agile practices | Organizational characteristic needed | NA | PA | LA | FA |
|---|---|---|---|---|---|
| Reflect and tune | ….. | | | | |
| **Collaborative planning** | Transparency of management | | | X | |
| | Small power-distance in the organization | | | | X |
| | Developers buy-in | | | | X |
| | Collaborative management style | | X | | |
| | Management buy-in | X | | | |
| Coding standards | ….. | | | | |

**NA:** Not Achieved (0%-35%)    **PA:** Partially Achieved (35%-65%)
**LA:** Largely Achieved (65%-85%)    **FA:** Fully Achieved (85%-100%)

**Table 2. Organizational Assessment Results**

Each of these organizational characteristics is assessed using a number of different questions. Depending on the question, a manager or developer within the organization, or the assessor himself or herself answers it. The 5 Levels of Agility incorporate approximately 300 indicators to measure the various organizational characteristics related to agile practices and concepts [43].

The result of the organizational assessment stage is a table that depicts the extent to which each organizational characteristic is achieved (see Table 2). This format for displaying results is beneficial to executives and decision makers as it draws attention to the characteristics of the organization that can cause problems in adopting a practice. Resembling project level assessment, determining the highest agile level an organization is capable of achieving is dependent on the organization's readiness to adopt the practices in that agile level. If the organizational characteristics needed for a practice are found to be *not achieved* or only *partially achieved*, then this is an indication that the organization is not ready to adopt that practice. As a result, the highest level of agility the organization can reach becomes the level at which a necessary organizational characteristic is missing. For example, in Table 2 since *collaborative planning* is in Agile Level 1, and since two of the characteristics that it needs are deficient, the highest level of agility for that organization is Level 1.

**3.2.3 Stage 4: Reconciliation.** Following the organizational readiness assessment, the agile level achievable by the organization is known. Prior to that, Stage 2 had identified the agile level that the project aspires to adopt. Therefore, the final step, reconciliation, is necessary to determine the agile practices the project will adopt. During this phase the differences between the projects' target level and the organization's readiness level are resolved to determine the final set of agile practices that will be adopted/employed. Three different scenarios are possible during this stage:

- *Organization Readiness Level > Project Target Level:* No reconciliation is needed and all the practices within the project's agile level and below become the chosen agile practices for adoption. This is a rare case because the project environment is usually contained with the organization.

- *Organization Readiness Level = Project Target Level:* No reconciliation is needed and all the practices within the project's agile level and below become the chosen agile practices for adoption. This is the ideal case since the project is achieving 100% of its agile potential.

- *Organization Readiness Level < Project Target Level:* Reconciliation is necessary. As discussed below, the framework provides two options for reconciling this situation.

(1) The first option relies on the how ready and willing the organization is for changes and improvements. The results of the organizational assessment have identified exactly which characteristics are hindering the organization from reaching higher levels of agility (i.e. the project's target level). If changing any of these characteristics is within the control of the organization, then the organization can undertake the necessary steps to improve them. If all of the recommended changes have been successfully made, then the organization can support agile practices at the project's target level. Otherwise, the projects' target level must be lowered accordingly.

(2) The second option is suitable for organizations that are unwilling to invest time, effort or money towards change, and only wants to adopt those agile practices that are within their current capacity. In this case, it is recommended to adopt only the agile practices the organization is ready for. The obvious downside to this approach is that the project is restricted to operating at a lower level of agility than its potential.

This reconciliation stage helps the organization in realistically identifying the agile practices it can adopt. At the same time, if the organization is able and willing to improve, then this stage guides it as to where the improvements need to occur so that the project can operate at its full agile potential. Moreover, by utilizing this approach, the organization prepares itself sufficiently before starting the process of introducing agile practices into the development process.

The next section provides a brief overview of the feedback gathered from presentations of the Agile Adoption Framework to members of the agile community.

## 4. Quantitative Feedback about the Agile Adoption Framework

The Agile Adoption Framework was presented to 28 members of the agile community. The feedback was gathered during 90-minute personal visits to the participants (or a group of them) in which the framework was presented and then discussed. After the presentation, the participants filled out a survey eliciting their feedback. In this section the results of the participants' feedback are examined from two perspectives, the first being the role or position of the participant, and the second being their years of experience. Additionally the feedback for the 5 Levels of Agility is presented separately from that of the 4-Stage process, since they were gathered through separate questionnaires.

### 4.1. Results for the 5 Levels of Agility.

The questionnaire concerning the 5 Levels of Agility focused on gathering feedback about its comprehensiveness, practicality, necessity, as well as whether the practices were placed at appropriate levels. Figure 4 illustrates that, in general, the participants were mostly in agreement with regard to comprehensiveness, practicality and necessity. However, some variability is observed among the participants concerning relevance. The most prominent concern was the position of the agile practices within the levels. We conjecture that this is due to the fact that each participant has different experiences, depending on their role, years of experience and the projects in which they have been involved. As a result, each participant places a different priority on the use of practices as reflected in their experiences. These beneficial insights and feedback have led us to recognize the utility of, and need for, the flexibility to tailor the 5 Levels of Agility to fit experiences and perhaps business goals. When examining the results classified by role, it is important to note that agile coaches and consultants had more positive feedback, in general, than the other positions. The results from the comprehensiveness, practicality and necessity show that there is in need for structure and guidance on how to organize these agile practices and concepts – this is exactly what the 5 Levels of Agility is intended to provide.

### 4.2. Results for the 4- Stage Process.

Figure 5 shows the feedback obtained relative to the 4-Stage assessment process. The feedback focused on the understandability of the process, its practicality, necessity, completeness, and effectiveness. As compared to the feedback on the 5 Levels of Agility, the feedback on the 4-Stage process is even more encouraging. Note that the agreement level is proportional to the years of experience and the roles of the individuals: the more experience and direct involvement with agile adoption, the higher the agreement rating. All of the highly experienced people strongly agreed that the process is clear and easy to understand. This can be expected, because the process is designed to model their particular activities. The completeness of the 4-Stage process had the lowest agreement percentage when compared to the other aspects of the process. We conjecture that a major factor contributing to this was the process used to gather the feedback. More specifically, only 90 minutes were allotted for presenting the framework to the participants, having follow-up discussions, and conducting the survey. We expect that this timeframe was too short for the participant (or anyone) to fully grasp the essence of the complete framework and the substantial set of relationships among its constituent components. This expectation is somewhat confirmed by the participants that returned the questionnaires at a later time (and having the time to reflect on the presentation and supporting material) – they both strongly agreed that the 4-Stage process is complete.

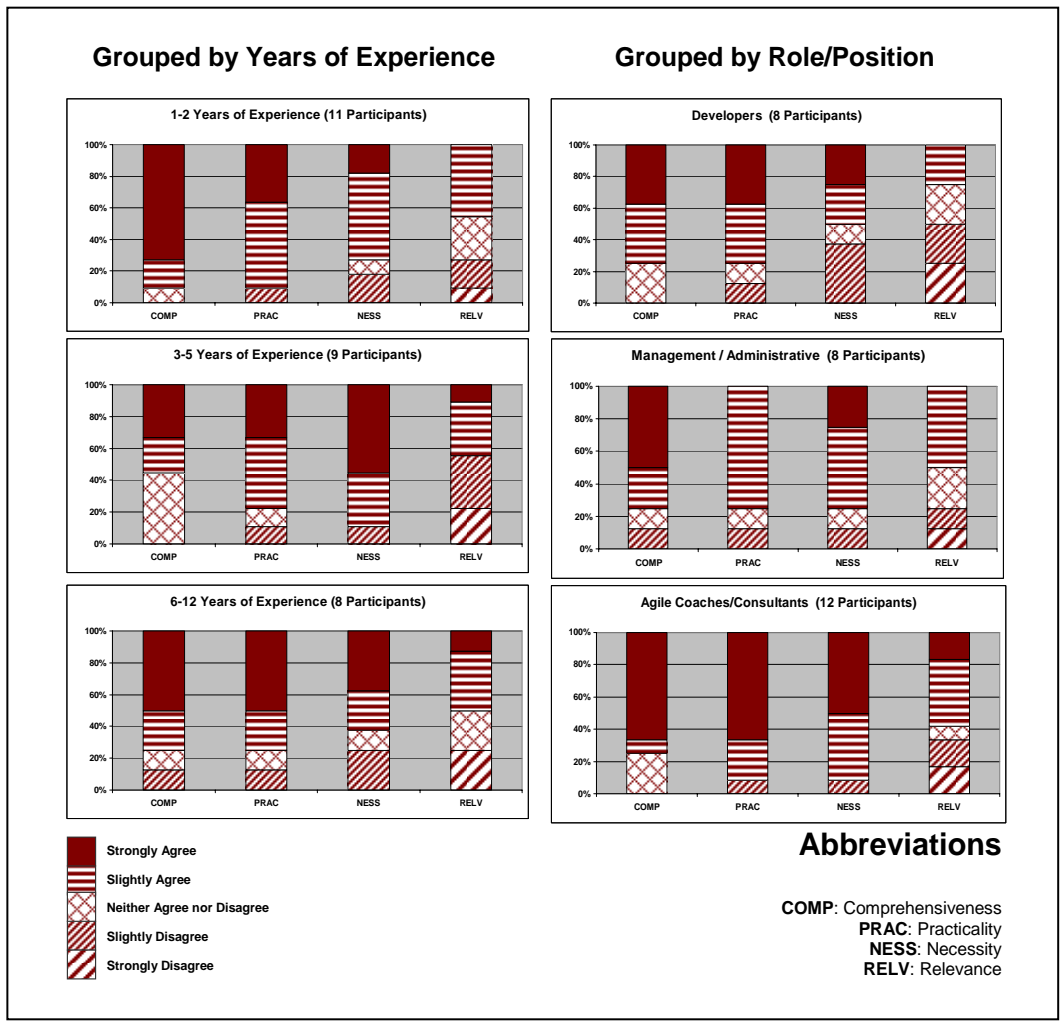

**Figure 4. Results of 5 Levels of Agility grouped by role and experience**

## 5. Conclusion

The Agile Adoption Framework is a first step toward addressing the need for providing organizations with a structured and repeatable approach to guide and assist them in the move toward agility. The framework is independent of any one particular agile method or style. Therefore, there are no restrictions on using XP or SCRUM or any other agile style within the framework. Moreover, the framework has two levels of assessment: one at the project level and another on an organizational level. Hence, it accommodates the uniqueness of each project, and at the same time, recognizes that each project is surrounded by, and is part of, an overall organization that must be ready to adopt the requisite agile practices. We view the Agile Adoption Framework as an initial contribution towards answering the complex question of *how* to adopt agile practices.

In summary, we propose this framework as an approach to guide and assist organizations in their quest to adopt agile practices. Through identifying and assessing the presence of *discontinuing factors*, organizations can make a go/no-go decision regarding the move toward agility. By determining the *target level for a project* and then assessing the *organization* to determine the extent to which it is ready to achieve that target level of agility, the framework manages to provide coaches with a realistic set of agile practices for the project to adopt. The 4-Stage process assessment, through its utilization of the 5 Levels of Agility, provides an extensive outline of the areas within the organization that need improvement *before* the adoption effort starts.

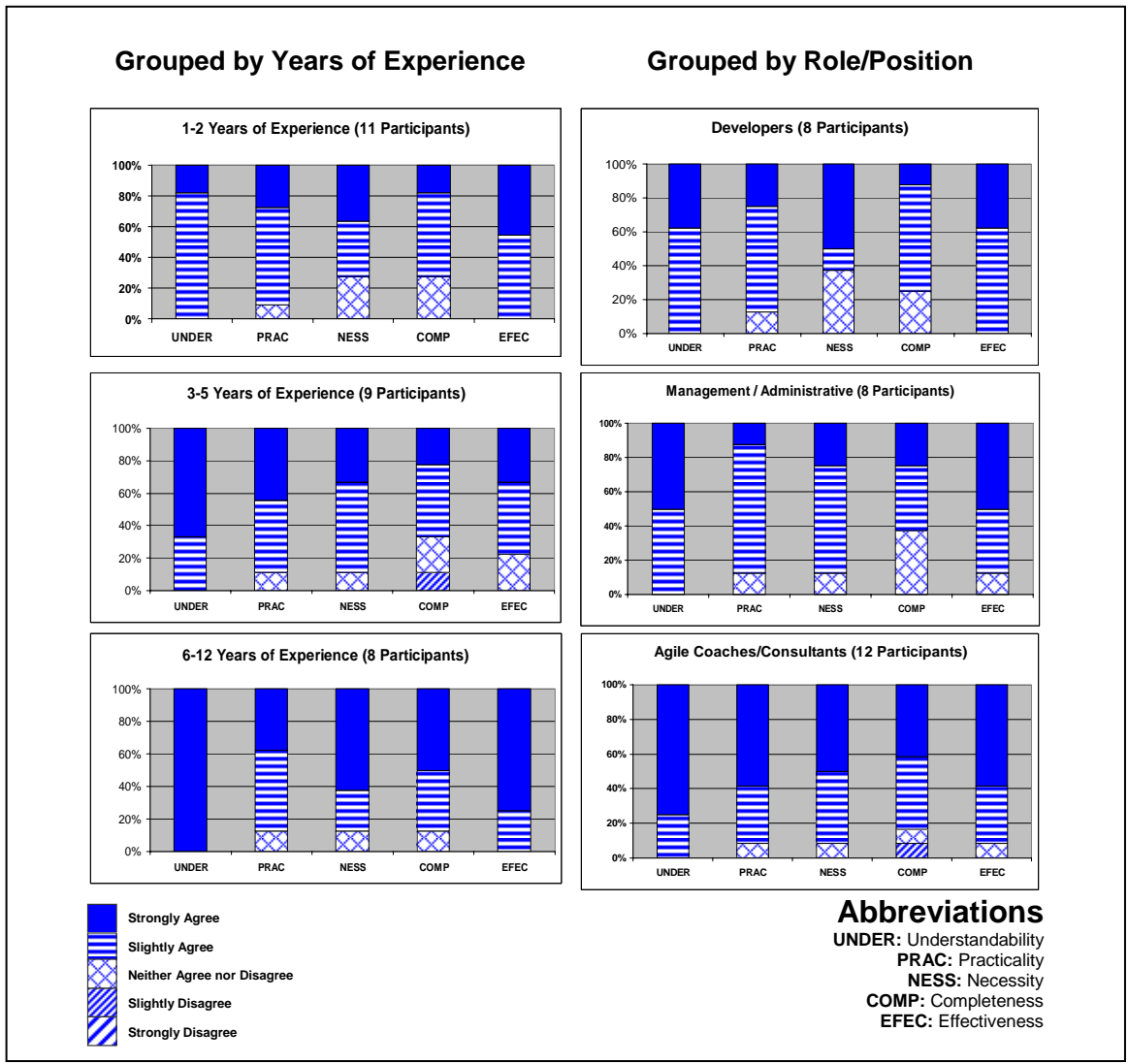

**Figure 5. Results of the 4-Stage Process grouped by role and experience**

While we recognize that the framework has significant room for improvement, we are encouraged by the comments given about the Agile Adoption Framework from members of the agile community:

- "*I think this is fantastic (work)*" –Agile consultant with 12 years experience

- "*This is the RIGHT time for this work! Excellent Job*" – Agile consultant with 8 years experience

- *"Overall this is first-class work and I endorse this work as legitimate in its interest and merit to our industry"* (paraphrased due to length*)* – XP Coach with 6 years experience